\documentclass[aps,prd,superscriptaddress,nofootinbib,tighten,preprint]{revtex4}

\include{header}

\usepackage{color}
\usepackage{amsfonts}
\usepackage{amssymb}
\usepackage{amsmath}
\usepackage{graphicx,subfigure}
\usepackage{bbm}
\def\slc#1{\setbox0=\hbox{$#1$}           
    \dimen0=\wd0                                 
    \setbox1=\hbox{/} \dimen1=\wd1               
    \ifdim\dimen0>\dimen1                        
       \rlap{\hbox to \dimen0{\hfil/\hfil}}      
       #1                                        
    \else                                        
       \rlap{\hbox to \dimen1{\hfil$#1$\hfil}}   
       /                                         
    \fi}

\begin{document}

\title{Bounds on Non-Standard Neutrino Interactions Using PINGU}

\author{Sandhya Choubey}
\email{sandhya@hri.res.in}

\affiliation{Department of Theoretical Physics, School of Engineering Sciences, KTH Royal Institute of Technology, AlbaNova University Center, 106 91 Stockholm, Sweden}
\affiliation{Harish-Chandra Research Institute, Chhatnag Road, Jhunsi, Allahabad 211 019, India}

\author{Tommy Ohlsson}

\email{tohlsson@kth.se}

\affiliation{Department of Theoretical Physics, School of Engineering Sciences, KTH Royal Institute of Technology, AlbaNova University Center, 106 91 Stockholm, Sweden}

\begin{abstract}
We investigate the impact of non-standard neutrino interactions (NSIs) on atmospheric neutrinos using the proposed PINGU experiment. In particular, we focus on the matter NSI parameters $\varepsilon_{\mu\tau}$ and $|\varepsilon_{\tau\tau} - \varepsilon_{\mu\mu}|$ that have previously been constrained by the Super-Kamiokande experiment. First, we present approximate analytical formulas for the difference of the muon neutrino survival probability with and without the above-mentioned NSI parameters. Second, we calculate the atmospheric neutrino events at PINGU in the energy range (2--100)~GeV, which follow the trend outlined on probability level. Finally, we perform a statistical analysis of PINGU. Using three years of data, we obtain bounds from PINGU given by $-0.0043~(-0.0048) < \varepsilon_{\mu\tau} < 0.0047~(0.0046)$ and $-0.03~(-0.016) < \varepsilon_{\tau\tau} < 0.017~(0.032)$ at 90~\% confidence level for normal (inverted) neutrino mass hierarchy, which improve the Super-Kamiokande bounds by one order of magnitude. In addition, we show the expected allowed contour region in the $\varepsilon_{\mu\tau}$-$\varepsilon_{\tau\tau}$ plane if NSIs exist in Nature and the result suggests that there is basically no correlation between $\varepsilon_{\mu\tau}$ and $\varepsilon_{\tau\tau}$.
\end{abstract}

\maketitle

\section{Introduction}

The phenomenon of neutrino oscillations has been found to be the leading description of neutrino flavor transitions. In the last decades, a vast amount of neutrino oscillation experiments have been performed and these experiments have been able to pin down the fundamental neutrino oscillation parameters to a good accuracy. However, despite the success of this description and due to the increasing precision of these experiments, there might be room for subleading new physics effects known as non-standard neutrino interactions (NSIs) \cite{Wolfenstein:1977ue,Valle:1987gv,Guzzo:1991hi,Roulet:1991sm,Grossman:1995wx}. Basically, NSIs give rise to dimension-six and higher-order operators, which lead to effective NSI parameters, and they arise in all modern extensions of the standard model (SM). It is important to note that both experimental and phenomenological bounds have been determined on such NSI parameters. In principle, NSIs can affect both production and detection processes and propagation in matter.

In this work, we will be interested in atmospheric neutrinos and therefore we will consider the matter NSI parameters $\varepsilon_{\mu\tau}$, $\varepsilon_{\mu\mu}$, and $\varepsilon_{\tau\tau}$ phenomenologically, which are important for neutrino oscillations in the $\nu_\mu$-$\nu_\tau$ sector. This means that we will set the other matter NSI para\-meters to zero, i.e.~$\varepsilon_{e\alpha} = 0$ for $\alpha = e, \mu, \tau$. Using the so-called two-flavor hybrid model and atmospheric neutrino data from the Super-Kamiokande I and II experiments, the following experimental bounds\footnote{Note the difference in the definition of the NSI parameters compared to the definition given in Ref.~\cite{Mitsuka:2011ty} by the Super-Kamiokande collaboration. This difference is described in detail in Ref.~\cite{Ohlsson:2012kf}.} have been obtain at 90~\% confidence level (C.L.) \cite{Mitsuka:2011ty}
$$
|\varepsilon_{\mu\tau}| < 0.033 \quad \mbox{and} \quad |\varepsilon_{\tau\tau} - \varepsilon_{\mu\mu}| < 0.147 \,.
$$
In addition, accelerator neutrino data from the MINOS experiment have set the bound $-0.200 < \varepsilon_{\mu\tau} < 0.070$  at 90~\% C.L.~\cite{Adamson:2013ovz}, which is less restrictive compared with the Super-Kamiokande bound on the same parameter.

Note that the corresponding model-independent phenomenological bounds on three matter NSI parameters for Earth matter are given by $|\varepsilon_{\mu\tau}| < 0.33$, $|\varepsilon_{\mu\mu}| < 0.068$, and $|\varepsilon_{\tau\tau}| < 21$ \cite{Biggio:2009nt}, where the last two bounds translate into a bound for $|\varepsilon_{\tau\tau} - \varepsilon_{\mu\mu}|$, which is less stringent than the above-mentioned experimental bound. This is particularly true for the bound on $|\varepsilon_{\tau\tau}|$. The main reason is that the model-independent phenomenological bound on $|\varepsilon_{\tau\tau}|$ comes from the one-loop contribution to the invisible $Z$-decay width at LEP, and is relatively lower compared to the bounds on other matter NSI parameters, since they are constrained by data from the charged lepton sector as well. On the other hand, the Super-Kamiokande bounds on the matter NSI parameters are indirect bounds coming from the role of the matter NSI parameters in neutrino oscillations and this is not necessarily poor for $|\varepsilon_{\tau\tau}|$. Therefore, the experimental bounds found by the Super-Kamiokande collaboration are the most stringent bounds on the three matter NSI parameters in the $\nu_\mu$-$\nu_\tau$ sector, which we will refer to in the rest of this work.

Outside the experimental collaborations, the role of matter NSI parameters in atmospheric neutrino experiments has been considered in a number of publications (see Refs.~\cite{Fornengo:2001pm,GonzalezGarcia:2004wg,Friedland:2004ah,Friedland:2005vy,GonzalezGarcia:2011my,Escrihuela:2011cf} and references therein). In particular, constraints on $\varepsilon_{\mu\tau}$ and $|\varepsilon_{\tau\tau} - \varepsilon_{\mu\mu}|$ were obtained from an analysis of the data from IceCube-79 and DeepCore \cite{Esmaili:2013fva}. In order to further investigate the discussed bounds on NSIs, we will use PINGU \cite{Koskinen:2011zz}, which is a proposed upgrade of the IceCube experiment at the South Pole. PINGU will be a huge experiment that will observe atmospheric neutrinos and probe high energies. It has been designed to measure the neutrino mass hierarchy, but it should also be able to reveal effects of NSIs. In this work, we will therefore use PINGU to predict bounds on the NSI parameters. The neutrino oscillation probabilities and predicted muon event rates from atmospheric neutrinos at PINGU in the energy range between 1~GeV and 20~GeV have been discussed in Ref.~\cite{Ohlsson:2013epa}. The impact of NSI parameters on the muon neutrino survival probability is known to grow with the increase of the neutrino energy. Therefore, we expect better constraints at higher energies. However, the atmospheric neutrino fluxes go down as about $E^{-2.7}$, and thus reducing the statistical significance of the data at higher energies. PINGU being a multi-megaton class ice detector will have better statistics. Therefore, in this work, we extend the previous study performed in Ref.~\cite{Ohlsson:2013epa} by increasing the energy range up to 100~GeV. However, in the previous study all matter NSI parameters were considered (except for $\varepsilon_{\mu\mu}$), whereas here we focus on the matter NSI parameters $\varepsilon_{\mu\tau}$ and $\varepsilon_{\tau\tau}$. We consider detector effective volume and resolution functions as quoted in the PINGU Letter of Intent \cite{Aartsen:2014oha} and include systematic uncertainties in the atmospheric neutrino fluxes in our analysis. We determine improved upper bounds on some NSI parameters using three years of running of the PINGU atmospheric neutrino experiment as described in Ref.~\cite{Aartsen:2014oha}. 

This work is organized as follows: In Section~\ref{sec:differences}, we discuss in detail the differences between standard and non-standard neutrino oscillation probabilities that are crucial for the understanding of the importance of NSIs. We focus on the analytical form of the $\nu_\mu \to \nu_\mu$ survival probability and its corresponding matter NSI parameters $\varepsilon_{\mu\tau}$, $\varepsilon_{\mu\mu}$, and $\varepsilon_{\tau\tau}$. Then, in Section~\ref{sec:events}, we compute the difference in the number of muon events from atmospheric neutrinos with and without NSI parameters. Next, in Section~\ref{sec:analysis}, we perform a statistical analysis for the proposed PINGU experiment and determine improved upper bounds on the NSI parameters $\varepsilon_{\mu\tau}$ and $\varepsilon_{\tau\tau}$. Finally, in Section~\ref{sec:s&c}, we summarize our results and draw our conclusions.

\section{Neutrino Oscillation Probability Differences}
\label{sec:differences}

In the case of three-flavor neutrino oscillations in matter, the expressions for ordinary neutrino oscillation probabilities are lengthy, and including NSIs, the expressions become even lengthier. In the standard framework, the neutrino oscillation probabilities are functions of eight parameters ($L$, $E$; $\theta_{12}$, $\theta_{13}$, $\theta_{23}$, $\delta$, $\Delta m_{21}^2$, and $\Delta m_{31}^2$) (where $L$ is the neutrino baseline length and $E$ is the neutrino energy), whereas for matter NSIs, they will be depending on six additional parameters. In general, differences between standard and non-standard neutrino oscillation probabilities will therefore depend on 14 parameters, i.e.~we have
\begin{equation}
\Delta P_{\alpha\beta} \equiv P_{\alpha\beta}^{\rm NSI} - P_{\alpha\beta}^{\rm SD} = \Delta P_{\alpha\beta} (L, E; \theta_{12}, \theta_{13}, \theta_{23}, \delta, \Delta m_{21}^2, \Delta m_{31}^2, \varepsilon_{ee}, \varepsilon_{e\mu}, \varepsilon_{e\tau}, \varepsilon_{\mu\mu}, \varepsilon_{\mu\tau}, \varepsilon_{\tau\tau}) \,.
\end{equation}

Expanding the $\nu_\mu \to \nu_\mu$ survival probability difference $\Delta P_{\mu\mu}$ to zeroth order in the parameter ratio $\Delta m^2_{21}/\Delta m^2_{31}$ and the mixing parameter $\sin \theta_{13}$ as well as to first order in the NSI parameters, one obtains \cite{Kopp:2007ne,Ribeiro:2007ud,Kikuchi:2008vq}
\begin{eqnarray}
\Delta P_{\mu\mu} &\simeq& - |\varepsilon_{\mu\tau}| c_{\phi_{\mu\tau}} A \left[ \sin^3 (2\theta_{23}) \Delta \sin (\Delta) + 4 \sin (2\theta_{23}) \cos^2 (2\theta_{23}) \sin^2 (\Delta/2) \right] \nonumber\\
&+& (|\varepsilon_{\mu\mu}| - |\varepsilon_{\tau\tau}|) A \sin^2 (2\theta_{23}) \cos(2\theta_{13}) \left[ \Delta \sin(\Delta)/2 - 2 \sin^2 (\Delta/2)\right] \,,
\label{eq:deltaPmm_approx}
\end{eqnarray}
where $\Delta \equiv \Delta m^2_{31} L/(2E) \equiv \Delta_{31} L$ and $A \equiv \sqrt{2} G_F N_e/\Delta_{31}$ with $G_F$ and $N_e$ being the Fermi coupling constant and the electron number
density in matter, respectively. Note that Eq.~(\ref{eq:deltaPmm_approx}) is only depending on the NSI parameters $\varepsilon_{\mu\tau}$, $\varepsilon_{\mu\mu}$, and $\varepsilon_{\tau\tau}$, and not on $\varepsilon_{ee}$, $\varepsilon_{e\mu}$, and $\varepsilon_{e\tau}$.

Now, using the so-called two-flavor hybrid model, which is an exact two-flavor neutrino oscillation model including matter NSIs with two NSI parameters $\varepsilon \equiv \varepsilon_{\mu\tau}$ and $\varepsilon' \equiv \varepsilon_{\tau\tau} - \varepsilon_{\mu\mu}$ \cite{Mitsuka:2011ty} and defining the auxiliary parameter
\begin{equation}
\Xi \equiv 1 + 2 A \cos (2\theta) \varepsilon' + A^2 \varepsilon'^2 \,,
\label{eq:Xi}
\end{equation}
we can perform a series expansion of the $\nu_\mu \to \nu_\mu$ survival probability to first order in the NSI parameter $\varepsilon$ in a similar way as for the three-flavor case, and we find that
\begin{eqnarray}
P_{\mu\mu} &=& 1 - \sin^2(2\theta) \frac{\sin^2 (\Delta_{2\nu} \sqrt{\Xi}/2)}{\Xi} \nonumber\\
&-& \varepsilon A \left\{ \sin^3 (2\theta) \frac{\Delta_{2\nu} \sqrt{\Xi} \sin(\Delta_{2\nu} \sqrt{\Xi})}{\Xi^2}  + 2 \sin(2\theta) \frac{[\cos (4\theta) - 1 + 2 \Xi] \sin^2 (\Delta_{2\nu} \sqrt{\Xi}/2)}{\Xi^2} \right\} \nonumber\\
&+& {\cal O}(\varepsilon^2) \,,
\label{eq:Pmm_hybrid_e}
\end{eqnarray}
where $\Delta_{2\nu} \equiv \Delta m^2 L/(2E)$.
Note that Eq.~(\ref{eq:Pmm_hybrid_e}) is exact in the other NSI parameter $\varepsilon'$.

Then, inserting Eq.~(\ref{eq:Xi}) into Eq.~(\ref{eq:Pmm_hybrid_e}) and making a series expansion to first order in $\varepsilon'$, we obtain the probability difference
\begin{eqnarray}
\Delta P_{\mu\mu} &=& - \varepsilon A [\sin^3 (2\theta) \Delta_{2\nu} \sin (\Delta_{2\nu}) + 4 \sin (2\theta) \cos^2 (2\theta) \sin^2 (\Delta_{2\nu}/2)] \nonumber\\
&-& \varepsilon' A \sin^2 (2\theta) \cos (2\theta) [\Delta_{2\nu} \sin (\Delta_{2\nu})/2 - 2 \sin^2 (\Delta_{2\nu}/2)]  + {\cal O}(\varepsilon^2,\varepsilon'^2) \,,
\label{eq:Pmm_hybrid_eep}
\end{eqnarray}
which by the replacements $\theta \to \theta_{23}$ and $\Delta_{2\nu} \to \Delta$ basically coincides with the corresponding three-flavor probability difference given in Eq.~(\ref{eq:deltaPmm_approx}). This means that the series expansion of the probability difference in the two-flavor hybrid model in both NSI parameters is an excellent approximation for the series expansion of the three-flavor probability difference in the three NSI parameters $\varepsilon_{\mu\tau}$, $\varepsilon_{\mu\mu}$, and $\varepsilon_{\tau\tau}$.

\begin{figure}[t]
\begin{center}
\includegraphics[width=0.495\textwidth]{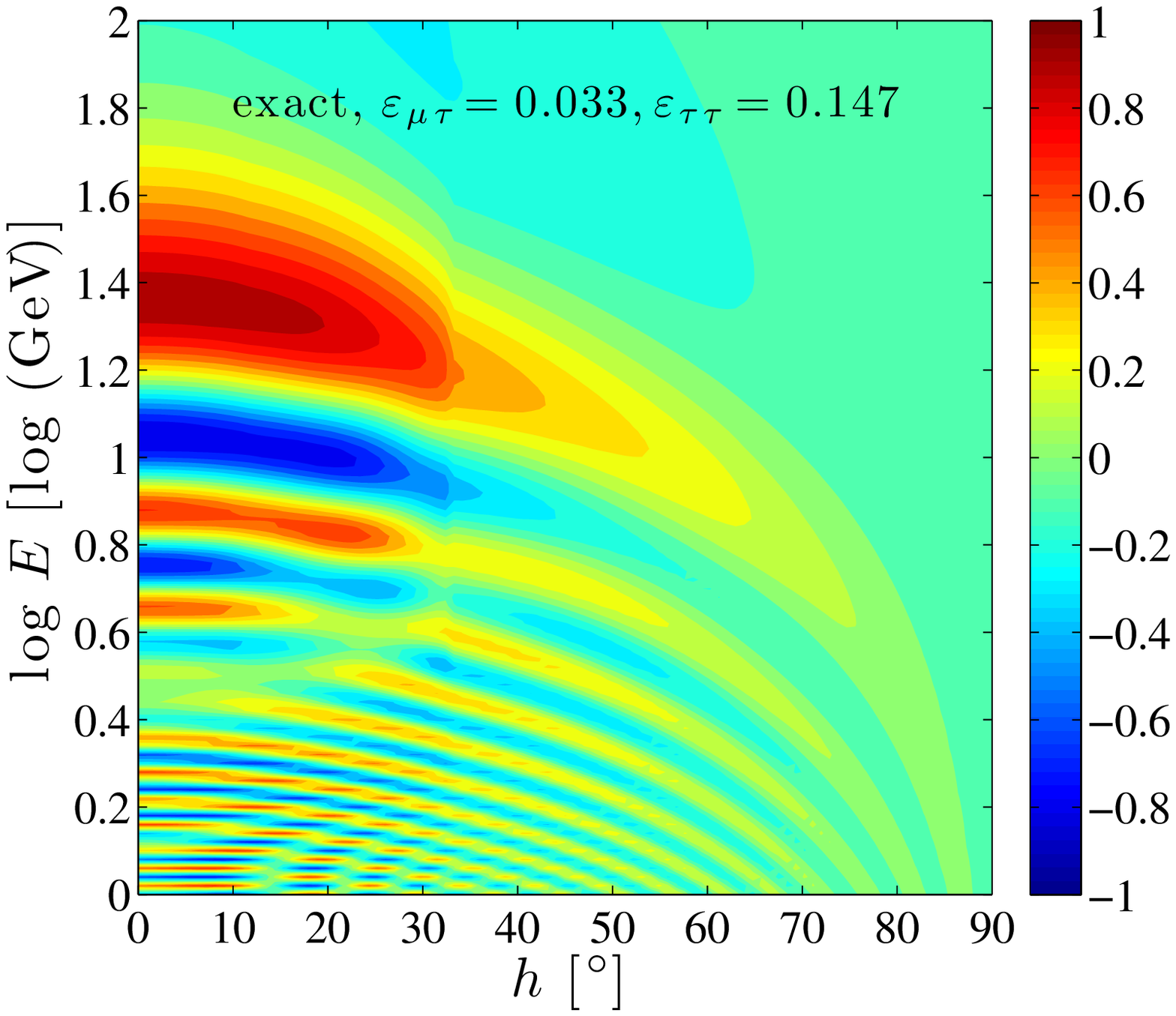}
\includegraphics[width=0.495\textwidth]{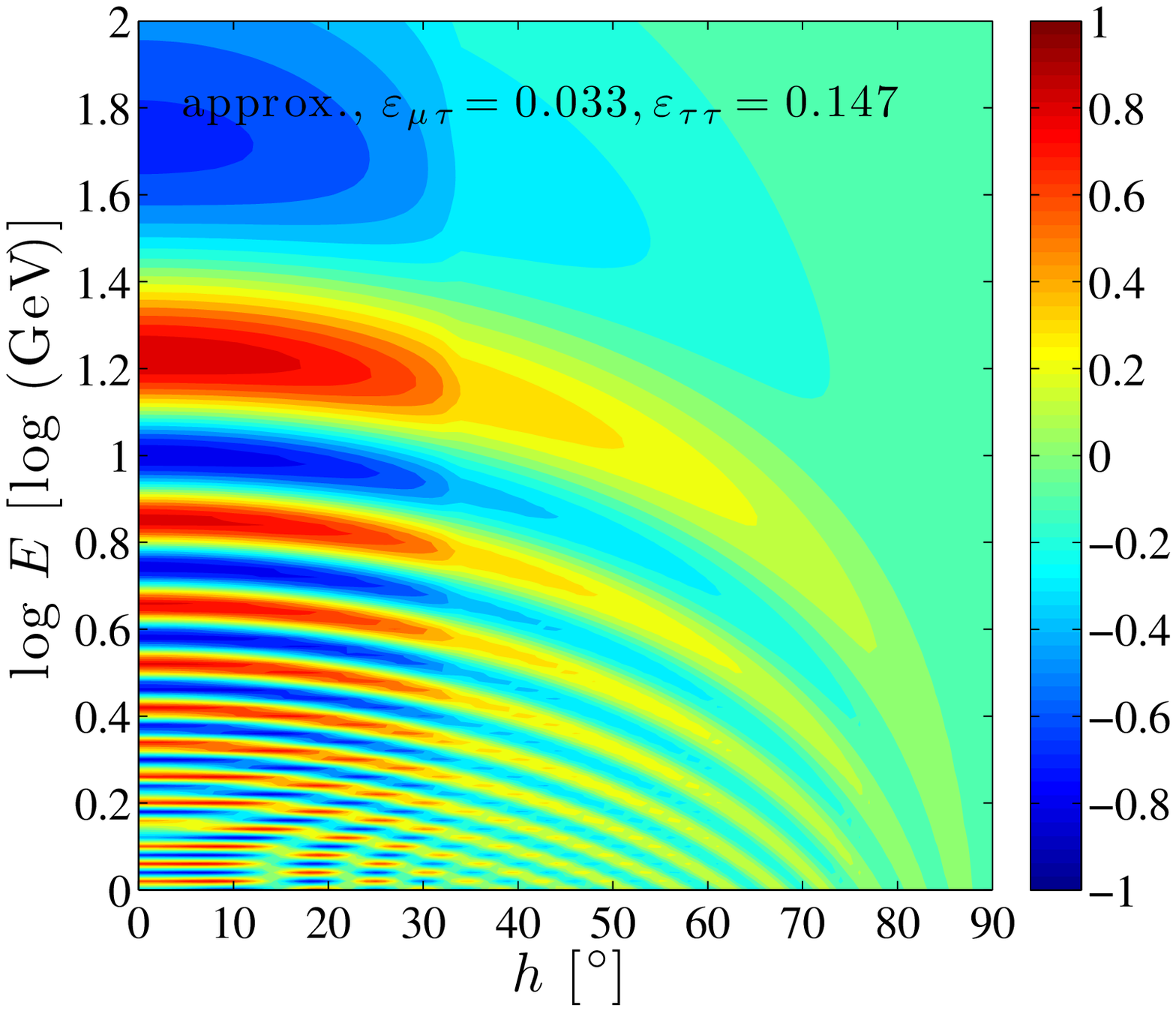}
\includegraphics[width=0.495\textwidth]{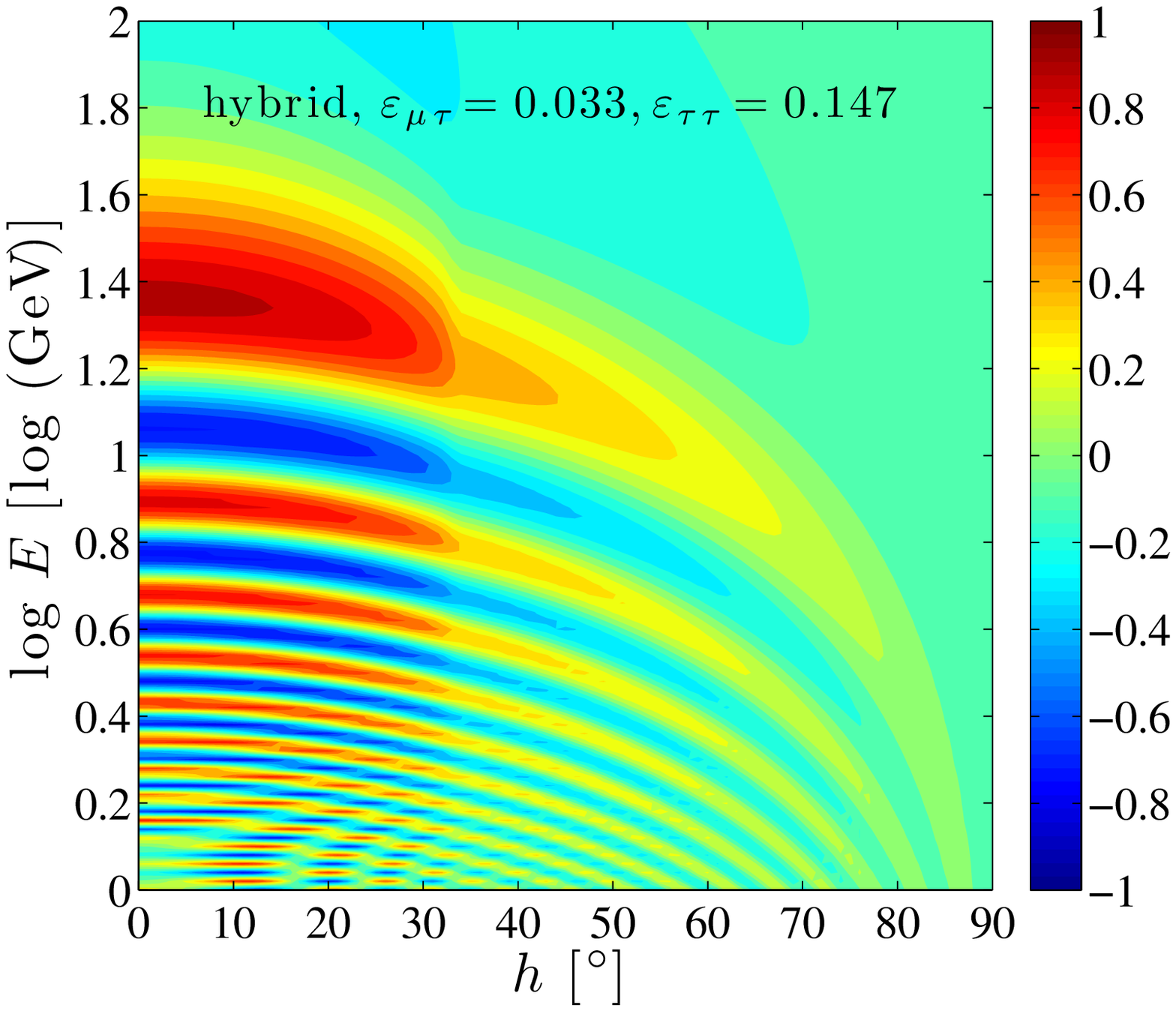}
\caption{The probability difference as a function of the nadir angle and the neutrino energy for three different cases: the exact three-flavor expression (upper-left plot), the approximative three-flavor expression (upper-right plot), and the exact two-flavor hybrid model expression (lower plot). The following values of the neutrino parameters have been used: $\theta_{12} = 33.57^\circ$, $\theta_{13} = 8.73^\circ$, $\theta_{23} = 41.9^\circ$, $\delta = 0$ (no leptonic CP-violation), $\Delta m^2_{21} = 7.45 \times 10^{-5} \, {\rm eV}^2$, $\Delta m^2_{31} = +2.417 \times 10^{-3} \, {\rm eV}^2$ (normal neutrino mass hierarchy), and the non-zero NSI parameter values $\varepsilon_{\mu\tau} = 0.033$ and $\varepsilon_{\tau\tau} = 0.147$.
}
\label{fig:deltaPmm_SK}
\end{center}
\end{figure}

In Fig.~\ref{fig:deltaPmm_SK}, assuming the upper bounds on the two-flavor NSI parameters $\varepsilon = 0.033$ and $\varepsilon' = 0.147$ from the Super-Kamiokande experiment \cite{Ohlsson:2012kf,Mitsuka:2011ty}, we present the probability difference $\Delta P_{\mu\mu}$ as a function of the nadir angle for the Earth and the neutrino energy using three different expressions: (i) the exact three-flavor expression for the probability difference with the PREM profile \cite{Dziewonski:1981xy} of the Earth's matter density, (ii) the approximative three-flavor expression in Eq.~(\ref{eq:deltaPmm_approx}) with a step-function profile, and (iii) the exact two-flavor hybrid model expression with a step-function profile.

In the case of the exact three-flavor expression for the probability difference shown in the upper-left plot of Fig.~\ref{fig:deltaPmm_SK}, we observe that the probability difference is maximal in a large region in parameter space for nadir angles $h$ between 0 and around $20^\circ$, i.e.~when neutrinos are basically going through the whole Earth, and for neutrino energies $E$ around 25~GeV ($\log 25 \simeq 1.4$). Moreover, there are smaller regions for which the probability difference is also maximal. Theses regions have neutrino energies which are smaller than 25~GeV, and are centered around approximately 11~GeV, 8~GeV, 6~GeV, and 5~GeV. We can now compare the results in the upper-left plot of Fig.~\ref{fig:deltaPmm_SK} with the results in the upper-right and lower plots. First, in the case of the approximative three-flavor expression (upper-right plot), we find that there appears a large region with energy larger than 30~GeV, and in addition, there are several other areas with energies smaller than 15~GeV. Furthermore, note that the neutrino energy of the largest maximal region has decreased from 25~GeV to 15~GeV. Second, in the case of the two-flavor hybrid model expression (lower plot), we note that the largest maximal region remains at around 25~GeV, but similar areas with energies smaller than 25~GeV appear as in the case of the approximative three-flavor expression. The differences compared to the exact three-flavor expression are effects of the approximations as well as the results presented in the upper-left plot uses the PREM profile and not a step-function profile as in the cases of the results in the upper-right and lower plots. In conclusion, the two-flavor hybrid model expression of the probability difference is an excellent approximation of the corresponding exact three-flavor expression, at least for neutrino energies of the order of 15~GeV or higher.

For values of the two-flavor mixing angle $\theta$ close to $45^\circ$, the most contributing term in the series expansion of the two-flavor hybrid model probability difference displayed in Eq.~(\ref{eq:Pmm_hybrid_eep}) is $- \varepsilon A \sin^3 (2\theta) \Delta_{2\nu} \sin (\Delta_{2\nu})$, which means that the energies of the oscillation maxima are given by (cf.~the discussion in Ref.~\cite{Ohlsson:2013epa})
\begin{equation}
E_{\max,n} \sim \frac{\Delta m^2 L}{2n\pi} \,, \quad \mbox{where $n = 1,2,3,\ldots$.}
\end{equation}
Using $\Delta m^2 = |\Delta m^2_{31}|$ and $L = 12~742~{\rm km}$ (i.e.~the maximal neutrino baseline length in the Earth corresponding to the nadir angle $h = 0$), one obtains the positions of the oscillation maxima at
$E_{\max,1} \simeq 24.8 \, {\rm GeV}$, $E_{\max,2} \simeq 12.4 \, {\rm GeV}$, $E_{\max,3} \simeq 8.3 \, {\rm GeV}$, $E_{\max,4} \simeq 6.2 \, {\rm GeV}$, and $E_{\max,5} \simeq 5.0 \, {\rm GeV}$,
which are more or less comparable with the values discussed above for the exact three-flavor probability difference (see the upper-left plot of Fig.~\ref{fig:deltaPmm_SK}).

\begin{figure}[!t]
\begin{center}
\includegraphics[width=0.495\textwidth]{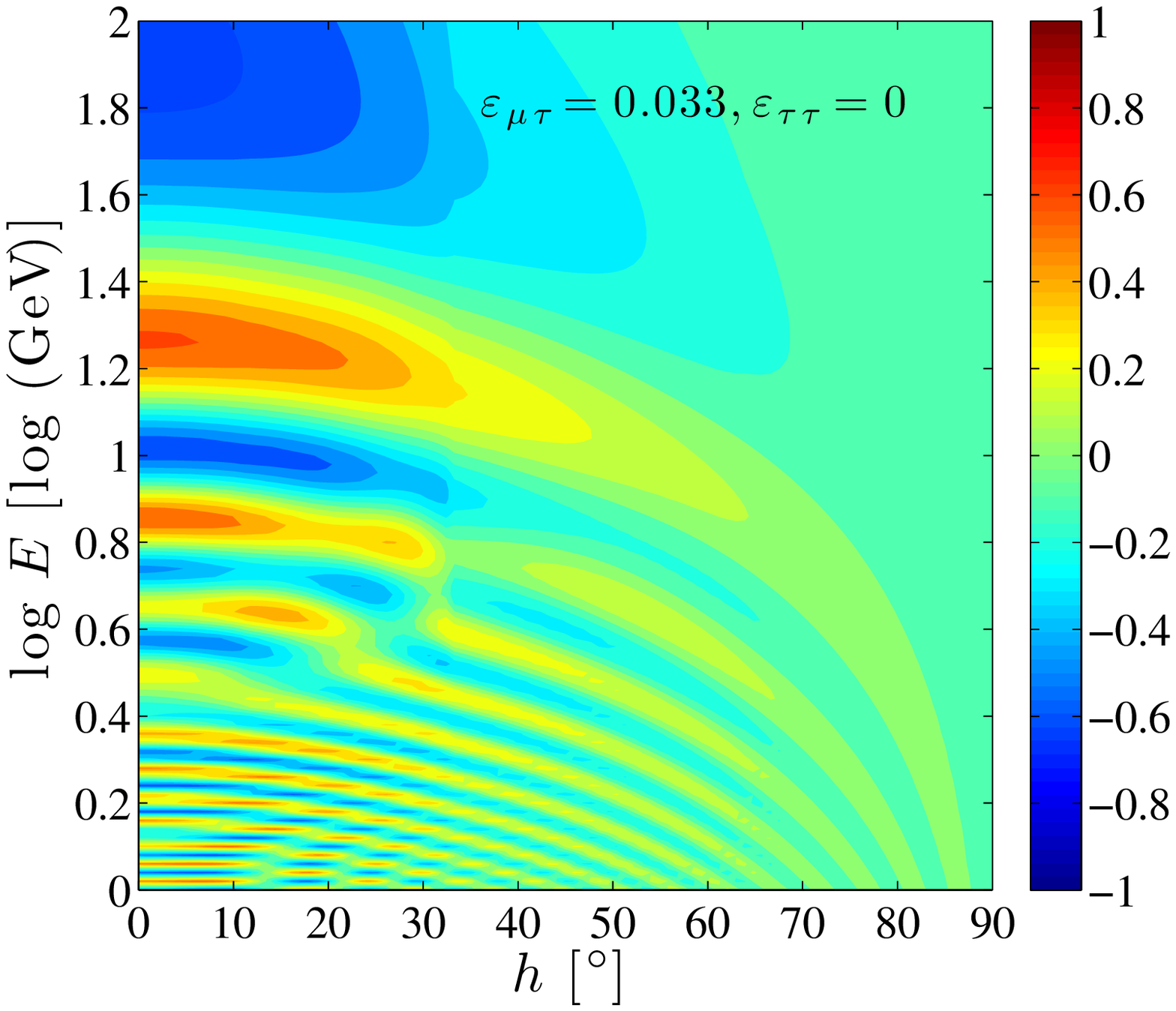}
\includegraphics[width=0.495\textwidth]{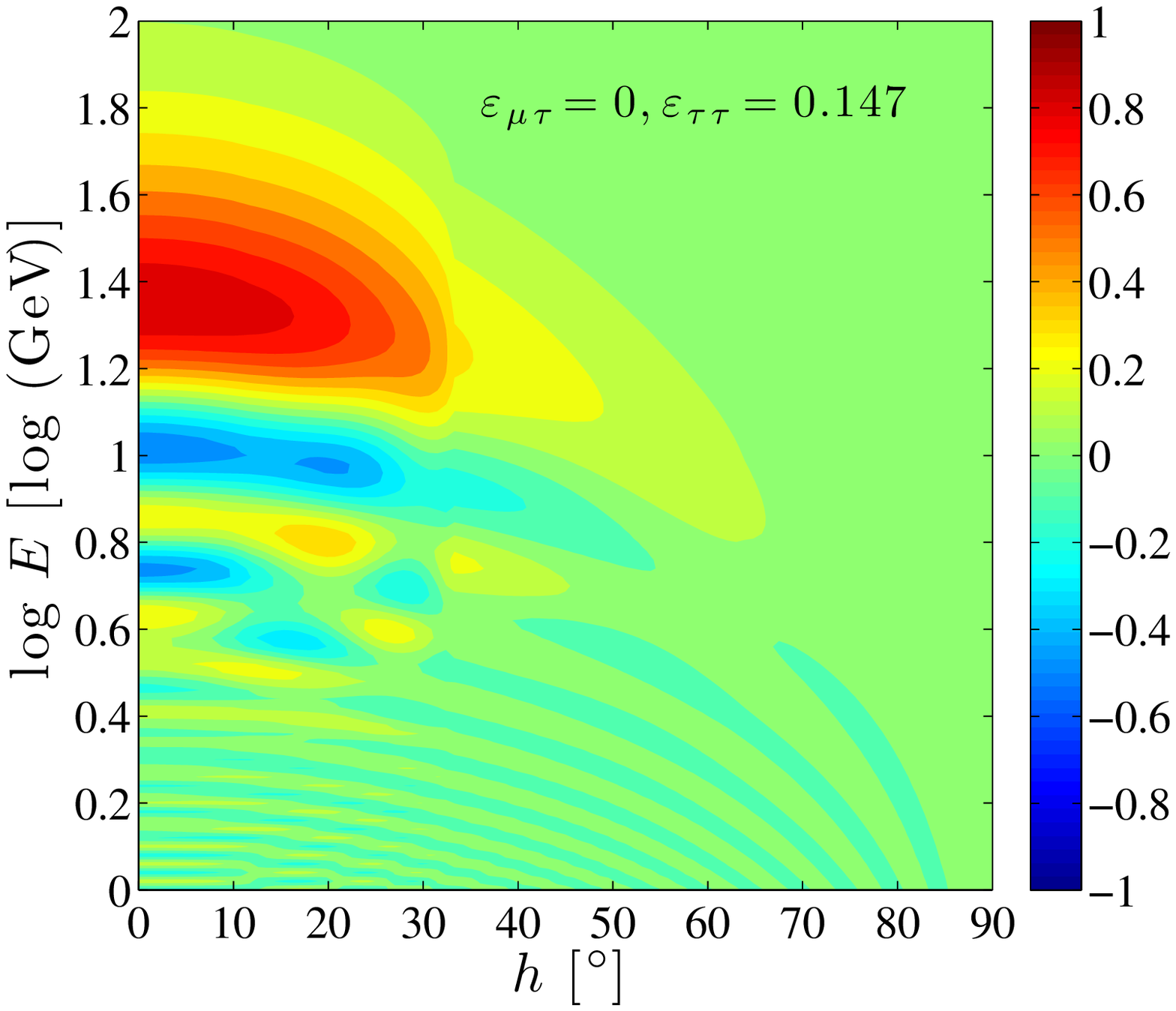}
\caption{The exact three-flavor probability difference as a function of the nadir angle and the neutrino energy for two different cases: $\varepsilon_{\mu\tau} = 0.033$ and $\varepsilon_{\tau\tau} = 0$ (left plot) as well as $\varepsilon_{\mu\tau} = 0$ and $\varepsilon_{\tau\tau} = 0.147$ (right plot). The other parameter values are the same as in Fig.~\ref{fig:deltaPmm_SK}.}
\label{fig:deltaPmm_SK_e_ep}
\end{center}
\end{figure}

In Fig.~\ref{fig:deltaPmm_SK_e_ep}, we present the exact three-flavor probability difference with $\varepsilon_{\mu\tau} = 0.033$ and $\varepsilon_{\tau\tau} = 0$ (left plot) as well as $\varepsilon_{\mu\tau} = 0$ and $\varepsilon_{\tau\tau} = 0.147$ (right plot). Note that we will set $\varepsilon_{\mu\mu} = 0$ throughout the rest of this work, which means that $\varepsilon' = \varepsilon_{\tau\tau}$, i.e.~the two-flavor parameter $\varepsilon'$ will be equivalent to the three-flavor parameter $\varepsilon_{\tau\tau}$. We observe that assuming a non-zero value for the NSI parameter $\varepsilon_{\mu\tau}$ the effects of the maximal region above 40~GeV is dominating (left plot), whereas assuming a non-zero value for the NSI parameter $\varepsilon_{\tau\tau}$ the effects of the maximal region centered around 25~GeV is the most pronounced (right plot).

\begin{figure}[!t]
\begin{center}
\includegraphics[width=0.495\textwidth]{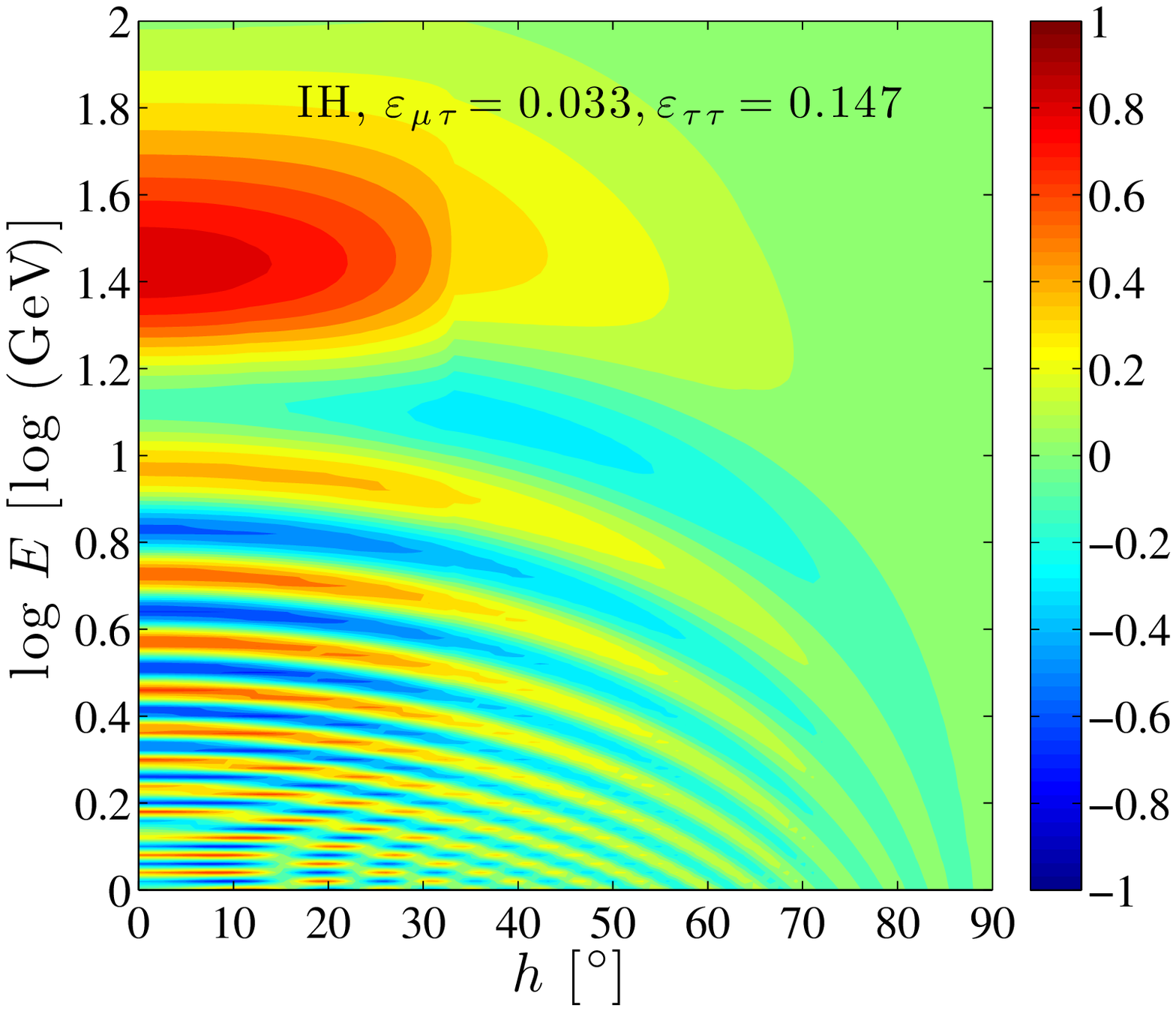}
\includegraphics[width=0.495\textwidth]{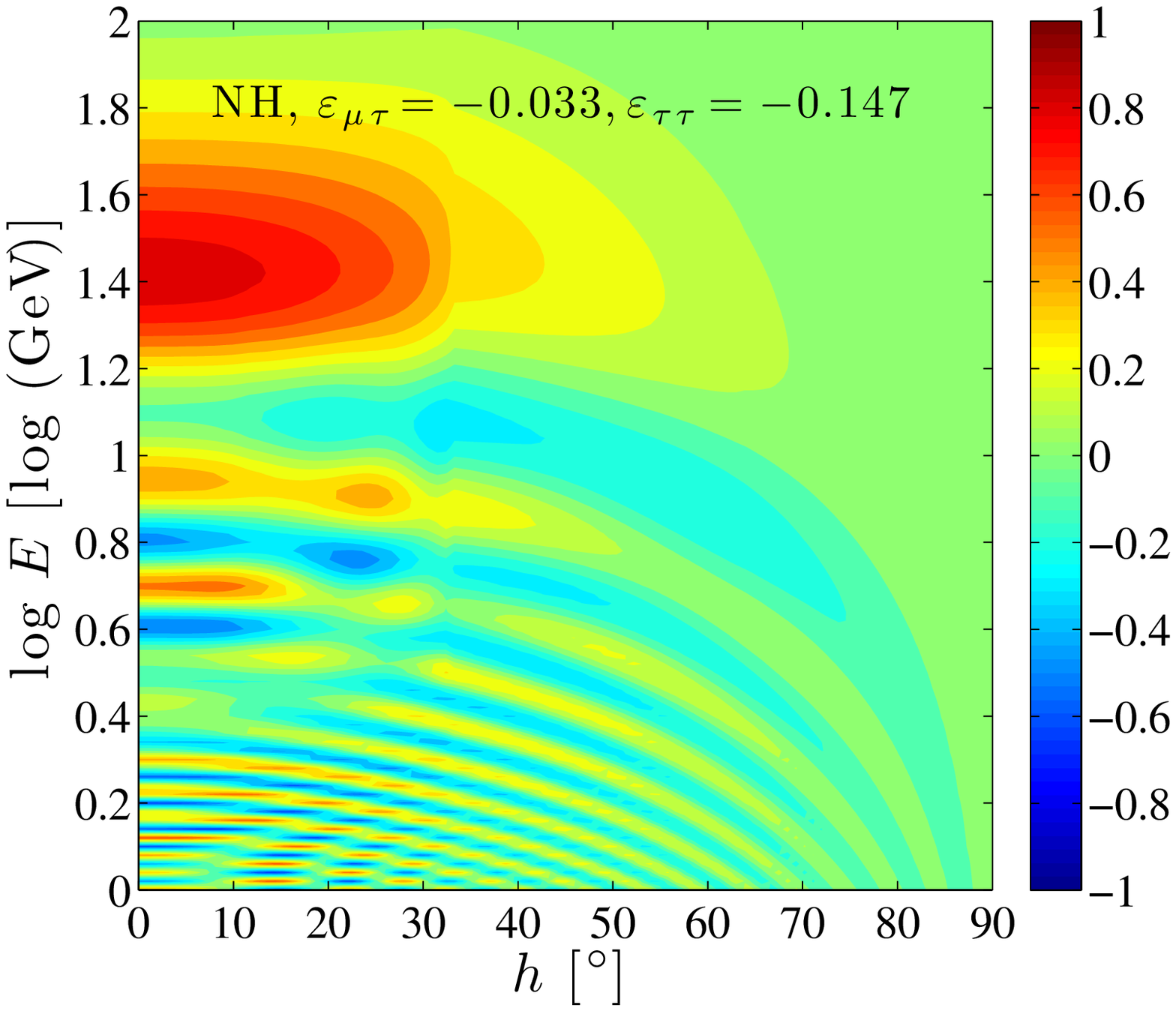}
\caption{The exact three-flavor probability difference as a function of the nadir angle and the neutrino energy for two different cases: $\varepsilon_{\mu\tau} = 0.033$ and $\varepsilon_{\tau\tau} = 0.147$ with inverted neutrino mass hierarchy (left plot) as well as $\varepsilon_{\mu\tau} = -0.033$ and $\varepsilon_{\tau\tau} = -0.147$ (right plot). The other parameter values are the same as in Fig.~\ref{fig:deltaPmm_SK}.}
\label{fig:deltaPmm_SK_IH_neg}
\end{center}
\end{figure}

Finally, in Fig.~\ref{fig:deltaPmm_SK_IH_neg}, we show the effects of assuming inverted neutrino mass hierarchy instead of normal neutrino mass hierarchy (left plot) and negative values of the NSI parameters instead of positive values (right plot). One can see that the results of choosing either the inverted neutrino mass hierarchy or negative values of the NSI parameters lead to similar results. Furthermore, the two plots should basically be compared to the upper-left plot of Fig.~\ref{fig:deltaPmm_SK}. Indeed, the effects of assuming the inverted neutrino mass hierarchy or negative values of the NSI parameters increase the maximal region of the probability difference around 30~GeV and the region is also slightly shifted to higher energies.

\section{Atmospheric Neutrino Events}
\label{sec:events}

The number of muon events from atmospheric neutrinos in the $i$th zenith bin and $j$th energy bin of a large underground ice/water Cherenkov detector such as PINGU is given by
\begin{align}
(N_\mu)_{ij}&= 2\pi \,N_T \,T \int_{(\cos\theta_z)_i}^{(\cos\theta_z)_{i+1}} {\rm d}\cos\theta_z
\int_{E_j}^{E_{j+1}} {\rm d}E \,
\int_{0}^{\pi} {\rm d}\cos\theta_z^\prime
\int_{0}^{\infty} {\rm d}E^\prime \,
\rho_{\rm ice}\,V_{\rm eff} 
R(E,E^\prime)R(\theta_z,\theta_{z^\prime}) \nonumber\\
&\times \left[\left(\frac{{\rm d}^2\phi_{\nu_\mu}}{{\rm d}\cos\theta^\prime_z {\rm d}E'}P_{\mu\mu} + 
\frac{{\rm d}^2\phi_{\nu_e}}{{\rm d}\cos\theta^\prime_z {\rm d}E'}P_{e\mu} \right)\,\sigma_{\nu} +
\left(\frac{{\rm d}^2\phi_{\bar\nu_\mu}}{{\rm d}\cos\theta^\prime_z {\rm d}E'}P_{\bar\mu\bar\mu} + 
\frac{{\rm d}^2\phi_{\bar\nu_e}}{{\rm d}\cos\theta^\prime_z {\rm d}E'}P_{\bar{e}\bar\mu} \right)\sigma_{\bar{\nu}}\right] \,,
\label{eq:evmu}
\end{align}
where $N_T$ is the number of targets per unit mass of the detector material, $T$ is the data taking time, and $\rho_{\rm ice}V_{\rm eff}$ is the effective volume of the detector, which is taken from the PINGU Letter of Intent \cite{Aartsen:2014oha} and the time $T$ is chosen as three years. The quantities, ${\rm d}^2\phi_{\nu_\mu}/{\rm d}\cos\theta^\prime_z {\rm d}E^\prime$ and ${\rm d}^2\phi_{\nu_e}/{\rm d}\cos\theta^\prime_z {\rm d}E^\prime$ are the differential atmospheric $\nu_\mu$ and $\nu_e$ fluxes, respectively, taken from Ref.~\cite{Honda:2004yz}, and $\sigma_\nu$ and $\sigma_{\bar\nu}$ are the interaction cross-sections for the neutrino and antineutrino with matter for which we use the following simplified formulas
\begin{align}
\sigma_\nu &= 7.3\times 10^{-39} \; E^\prime~{\rm m^2} \,, \\
\sigma_\nu &= 3.77\times 10^{-39} \; E^\prime~{\rm m^2} \,.
\end{align}
The distinction between primed and unprimed variables is as follows. The primed variables $E^\prime$ and $\theta_z^\prime$ are the true energy and zenith angle of the neutrinos, while the unprimed variables $E$ and $\theta_z$ are the corresponding measured or reconstructed ones.\footnote{The zenith and nadir angles are related to each other as $\theta_z + h = 180^\circ$.} The resolution functions $R(E,E^\prime)$ for energy and $R(\theta_z,\theta_{z^\prime})$ for zenith angle relate the true variables to the measured or reconstructed variables. We have taken Gaussian forms for the resolution functions 
\begin{equation}
R(p,p^\prime) = \frac{1}{\sqrt{2\pi}\sigma_p} \exp\left[\frac{(p-p^\prime)^2}{2\sigma_p^2}\right] \,.
\end{equation}
We have used $\sigma_E = 0.2 \; E^\prime$ and $\sigma_{\theta_z}= 0.5/\sqrt{E^\prime}$, which agree with the simulation results presented by the PINGU collaboration in Ref.~\cite{Aartsen:2014oha}. The flavor survival and conversion probabilities $P_{\mu\mu}$ and $P_{e\mu}$ for neutrinos and $P_{\bar\mu\bar\mu}$ and $P_{\bar e \bar\mu}$ for antineutrinos are calculated numerically by solving the three-generation differential equation of motion for neutrinos coming from the atmosphere and reaching the detector at the South Pole. This calculation uses the PREM profile \cite{Dziewonski:1981xy} for both standard oscillations and oscillations in the presence of NSI parameters. In the calculation, for the sake of completeness, we have also included the electron events in PINGU coming from atmospheric neutrinos. The number of electron events is given by an expression identical to Eq.~(\ref{eq:evmu}) in all respect with suitable changes. However, the electron events are not expected to make any significant contribution to the constraints on the NSI parameters $\varepsilon_{\mu\tau}$ and $\varepsilon_{\tau\tau}$. 

\begin{figure}[!t]
\begin{center}
\includegraphics[width=0.495\textwidth]{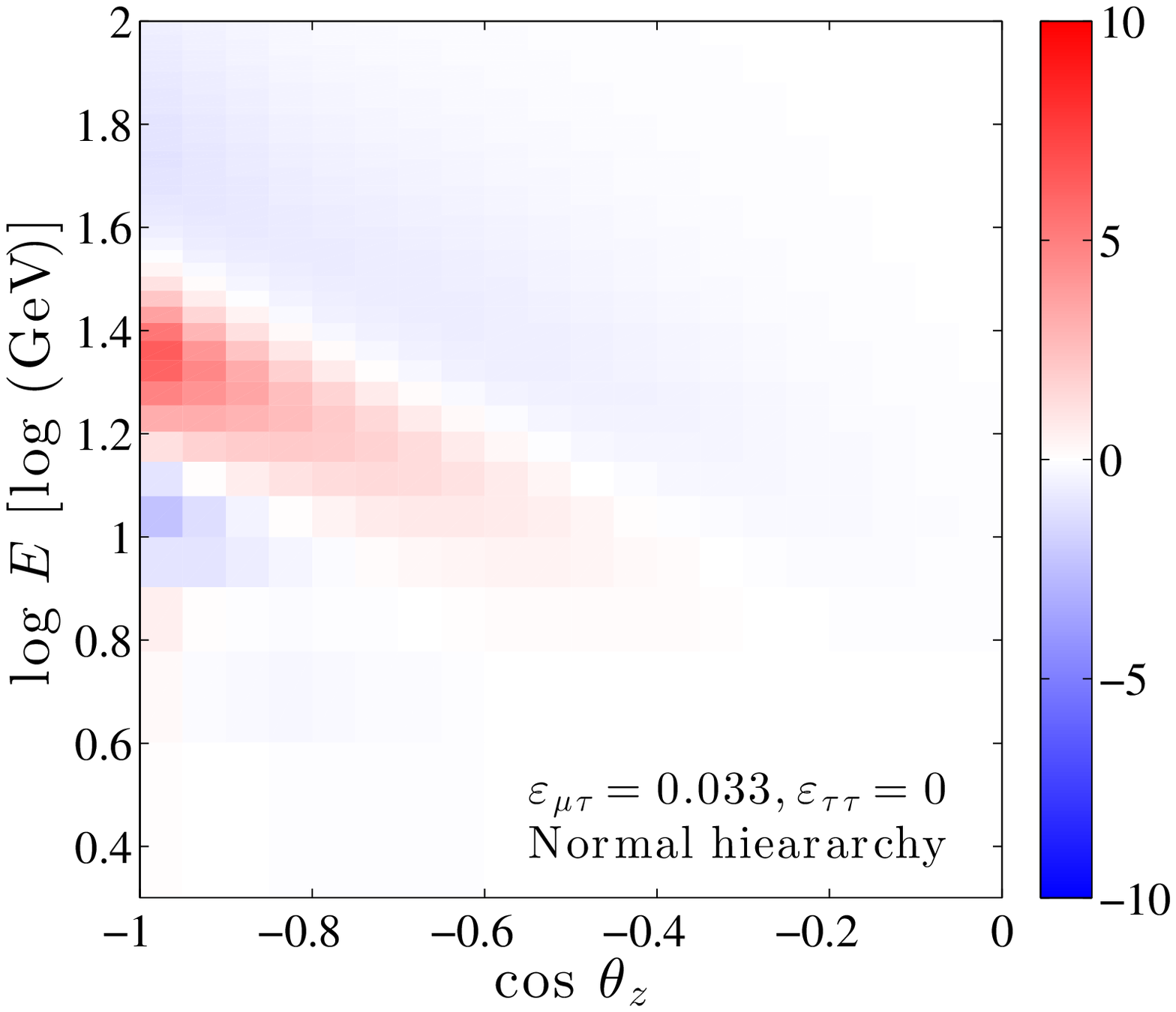}
\includegraphics[width=0.495\textwidth]{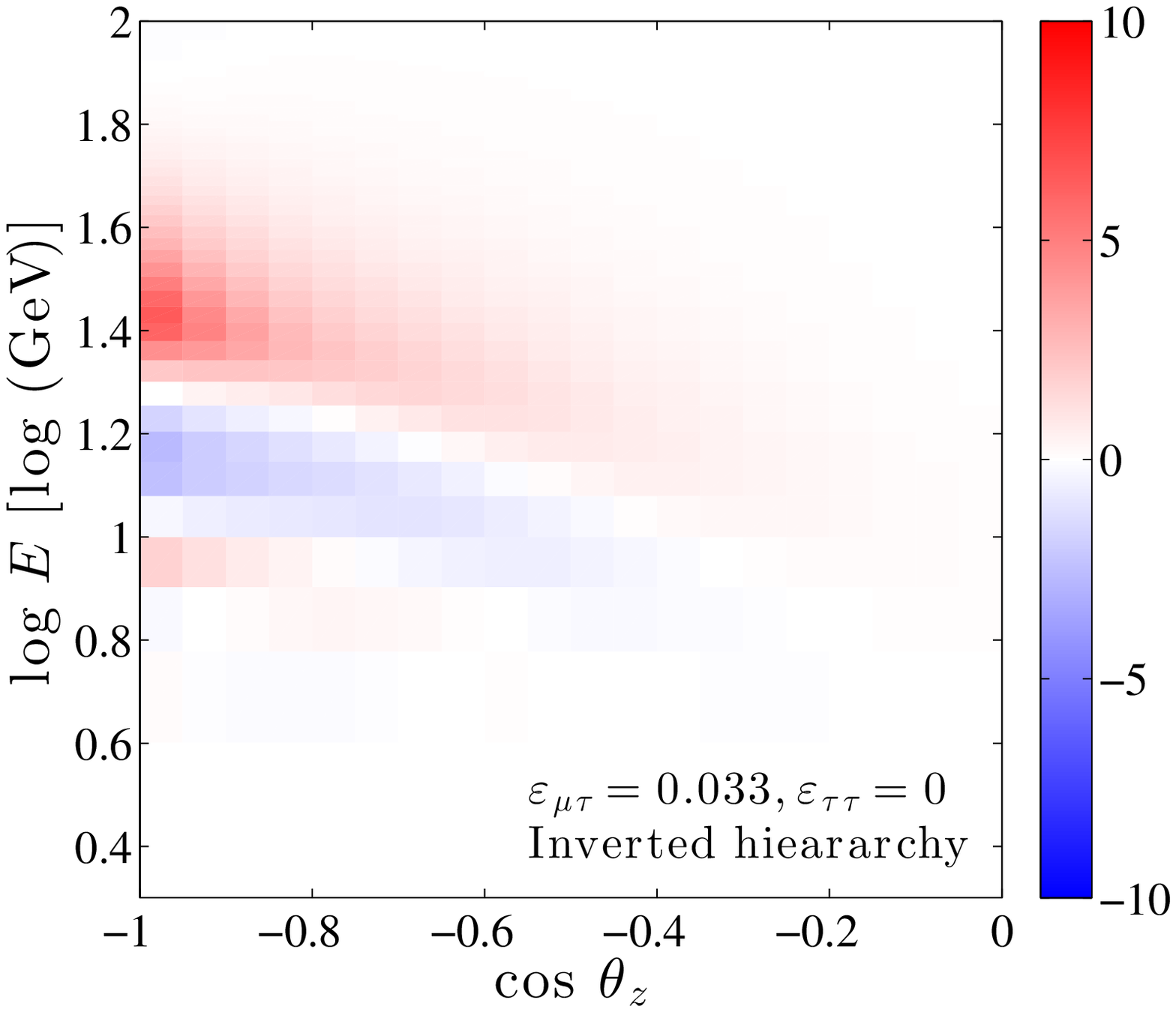}
\caption{The asymmetry in the number of muon events defined as $(N_{\rm NSI} - N_{\rm SM})/\sqrt{N_{\rm SM}}$ in the $\cos\theta_z$-${\rm log}E$ plane. The left plot is for normal neutrino mass hierarchy, whereas the right plot is for inverted hierarchy. The benchmark values of $\varepsilon_{\mu\tau}$ and $\varepsilon_{\tau\tau}$ assumed for the plots are shown.}
\label{fig:diffepsmt}
\end{center}
\end{figure}

\begin{figure}[!t]
\begin{center}
\includegraphics[width=0.495\textwidth]{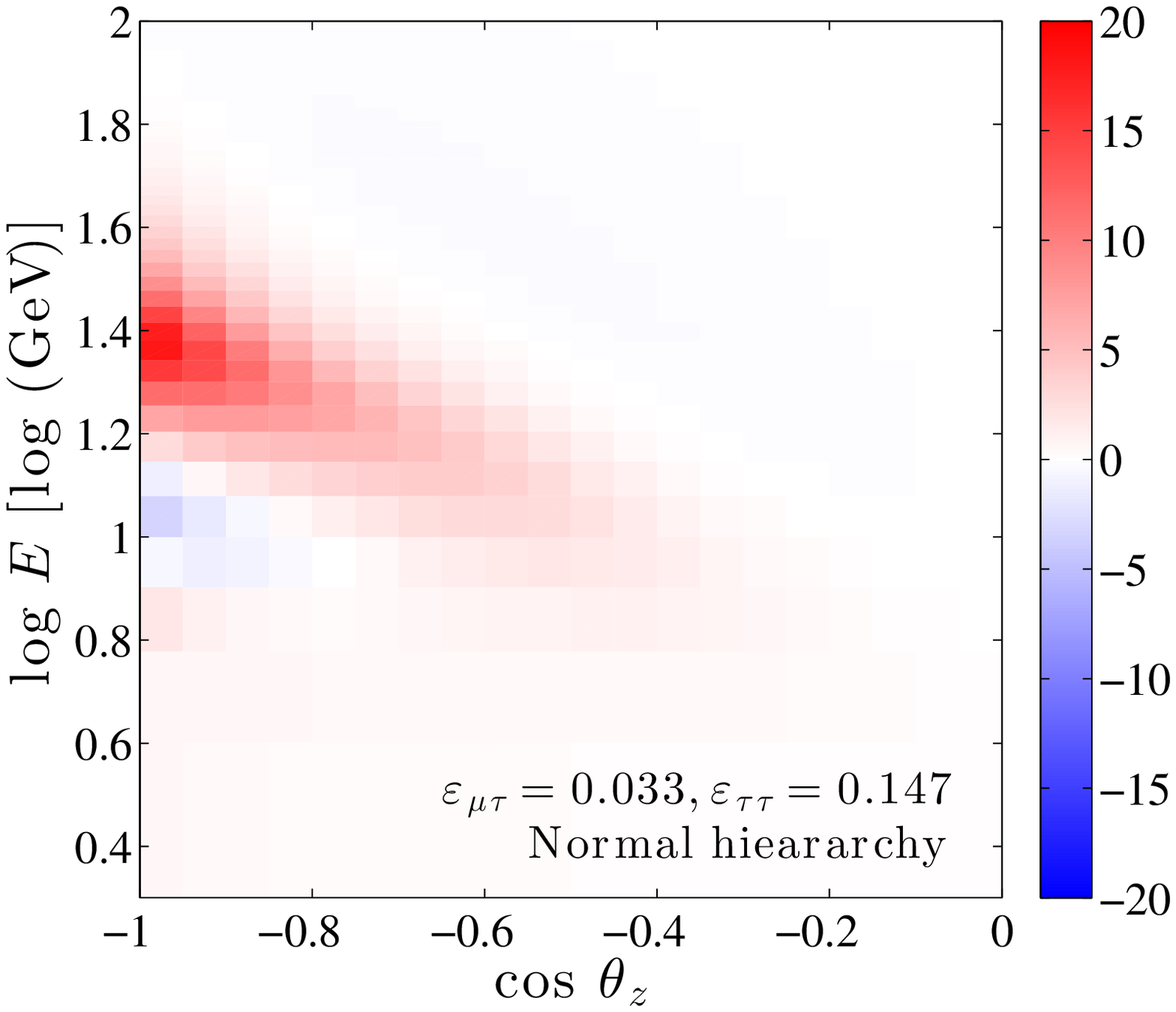}
\includegraphics[width=0.495\textwidth]{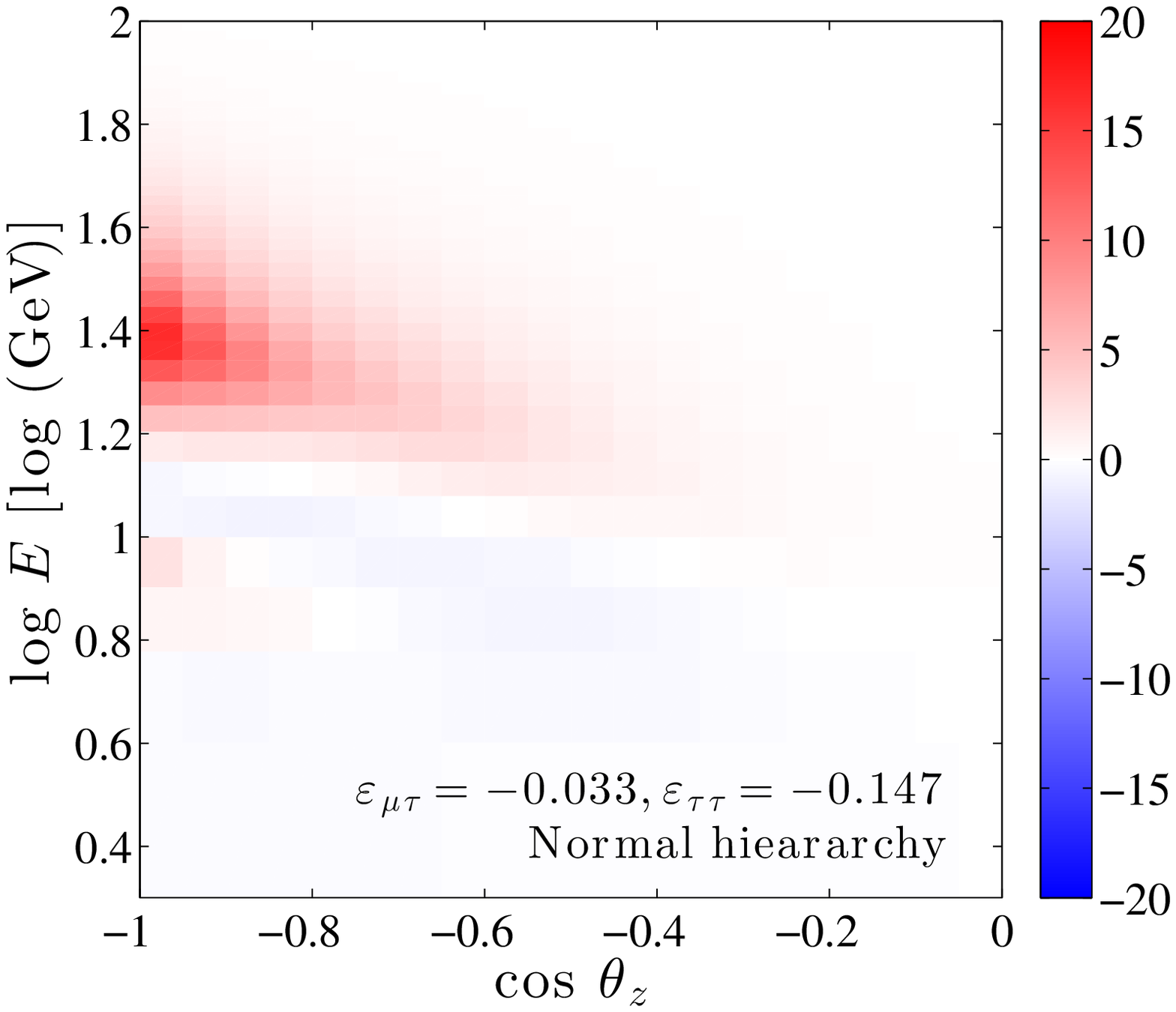}
\caption{The asymmetry in the number of muon events defined as $(N_{\rm NSI} - N_{\rm SM})/\sqrt{N_{\rm SM}}$ in the $\cos\theta_z$-${\rm log}E$ plane. The benchmark values of $\varepsilon_{\mu\tau}$ and $\varepsilon_{\tau\tau}$ assumed for the plots are shown. We have chosen normal neutrino mass hierarchy in both plots.}
\label{fig:diffepsboth}
\end{center}
\end{figure}

Since the neutrino oscillation probabilities depend on the presence and strength of the NSI parameters, as discussed in the previous section, we expect that to be reflected in the event spectrum observed at neutrino detectors. As we expect only the muon events to be significantly affected by the NSI parameters $\varepsilon_{\mu\tau}$ and $\varepsilon_{\tau\tau}$, we show the impact of these NSI parameters on the muon events in Figs.~\ref{fig:diffepsmt} and \ref{fig:diffepsboth}. We plot the quantity $(N_{\rm NSI} - N_{\rm SM})/\sqrt{N_{\rm SM}}$ in the $\cos\theta_z$-${\rm log}E$ plane, where $N_{\rm NSI}$ and $N_{\rm SM}$ are the numbers of expected muon type events in the presence and absence of additional NSIs, respectively. This gives the difference in the number of expected muon events at the detector, normalized to the square root of expected number muon events in the absence of any new physics. It can be interpreted as a rough estimate of the statistical significance with which the NSI parameters $\varepsilon_{\mu\tau}$ and $\varepsilon_{\tau\tau}$ can be constrained at PINGU, with three years of data. 

In Fig.~\ref{fig:diffepsmt}, we show this quantity for the case where $\varepsilon_{\mu\tau}=0.033$ and $\varepsilon_{\tau\tau}=0$, while in Fig.~\ref{fig:diffepsboth}, we allow both the NSI parameters to be non-zero and at their current Super-Kamiokande bounds at 90~\% C.L., i.e.~$|\varepsilon_{\mu\tau}|=0.033$ and $|\varepsilon_{\tau\tau}|=0.147$. The left (right) plot of Fig.~\ref{fig:diffepsmt} is for normal (inverted) neutrino mass hierarchy. A direct comparison of the left plot of this figure can be made with the left plot of Fig.~\ref{fig:deltaPmm_SK_e_ep}. The blue and red regions of the two plots are seen to be congruent with each other, where the additional smearing in the event plot is due to the finite detector resolutions involved. Nonetheless, for the core-crossing bins, one can clearly observe an excess of events due to NSIs at $E\simeq 25$~GeV, and a depletion at $E\simeq 10$~GeV and in the broad energy range between $E\in (40-100)$~GeV. Comparison of the right plot (for inverted hierarchy) of Fig.~\ref{fig:diffepsmt} with the left plot (for normal hierarchy) reveals that the distribution of the difference in the number of events due to NSIs changes significantly with the changing of the neutrino mass hierarchy. 

In Fig.~\ref{fig:diffepsboth}, we repeat this exercise for the case where both the NSI parameters $\varepsilon_{\mu\tau}$ and $\varepsilon_{\tau\tau}$ are non-zero simultaneously. The left plot is for $\varepsilon_{\mu\tau}=0.033$ and $\varepsilon_{\tau\tau}=0.147$, while the right plot shows the impact of NSIs with the sign of both $\varepsilon_{\mu\tau}$ and $\varepsilon_{\tau\tau}$ flipped. For both plots we have assumed normal neutrino mass hierarchy. Switching on $|\varepsilon_{\tau\tau}|$ has the main effect of increasing the muon neutrino survival probability over most of the energy range considered. As a result, the red regions in Fig.~\ref{fig:diffepsboth} become brighter compared to the ones in Fig.~\ref{fig:diffepsmt}, while the blue regions diminish and almost go away. From Fig.~\ref{fig:diffepsboth} one can also note that the sign of $\varepsilon_{\mu\tau}$ and $\varepsilon_{\tau\tau}$ does not make much of a difference to the expected event rates. This was also noted in Fig.~\ref{fig:deltaPmm_SK_IH_neg}.

\section{$\chi^2$ Analysis and Constraints on NSIs}
\label{sec:analysis}

\begin{figure}[t]
\begin{center}
\includegraphics[width=0.495\textwidth]{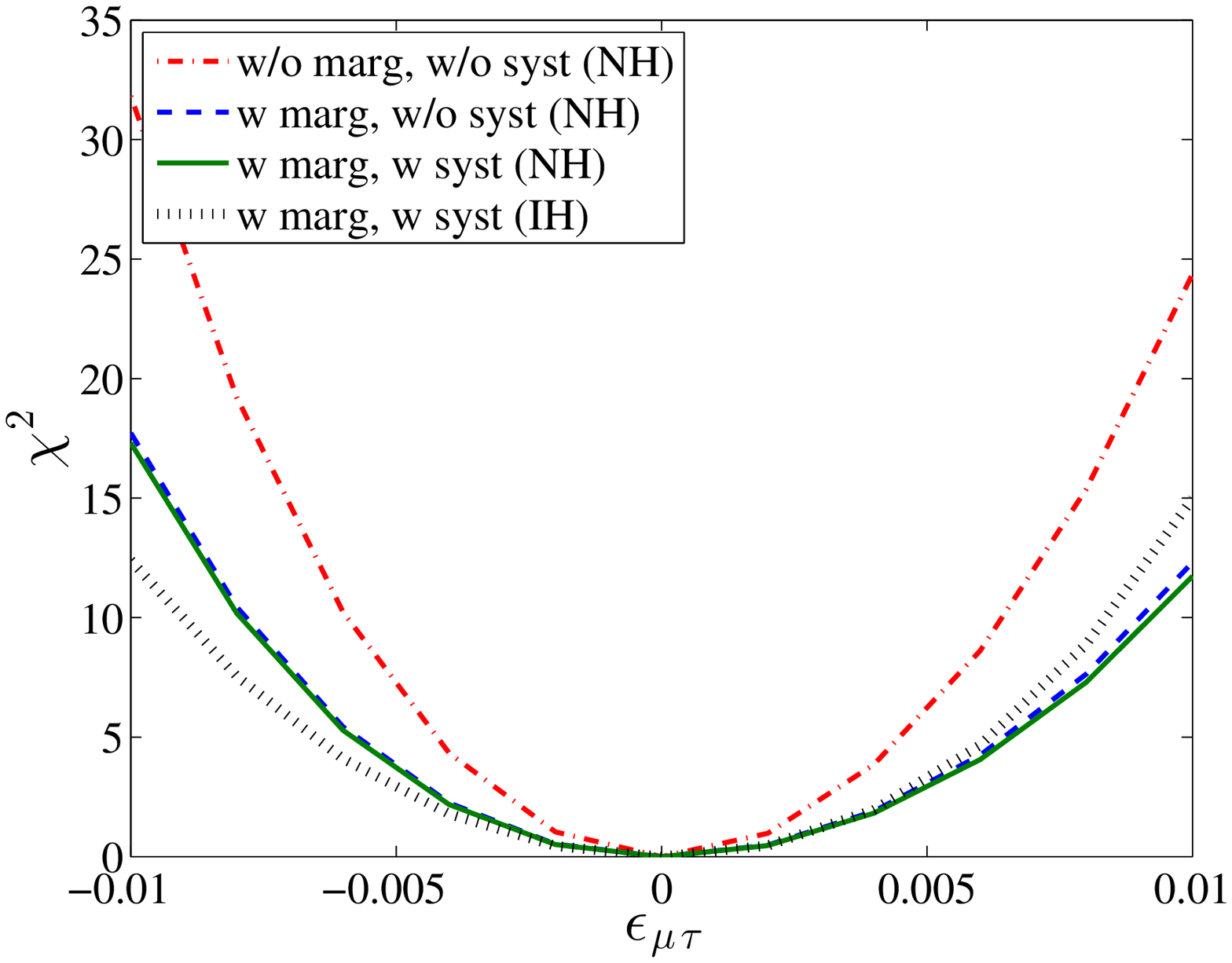}
\includegraphics[width=0.495\textwidth]{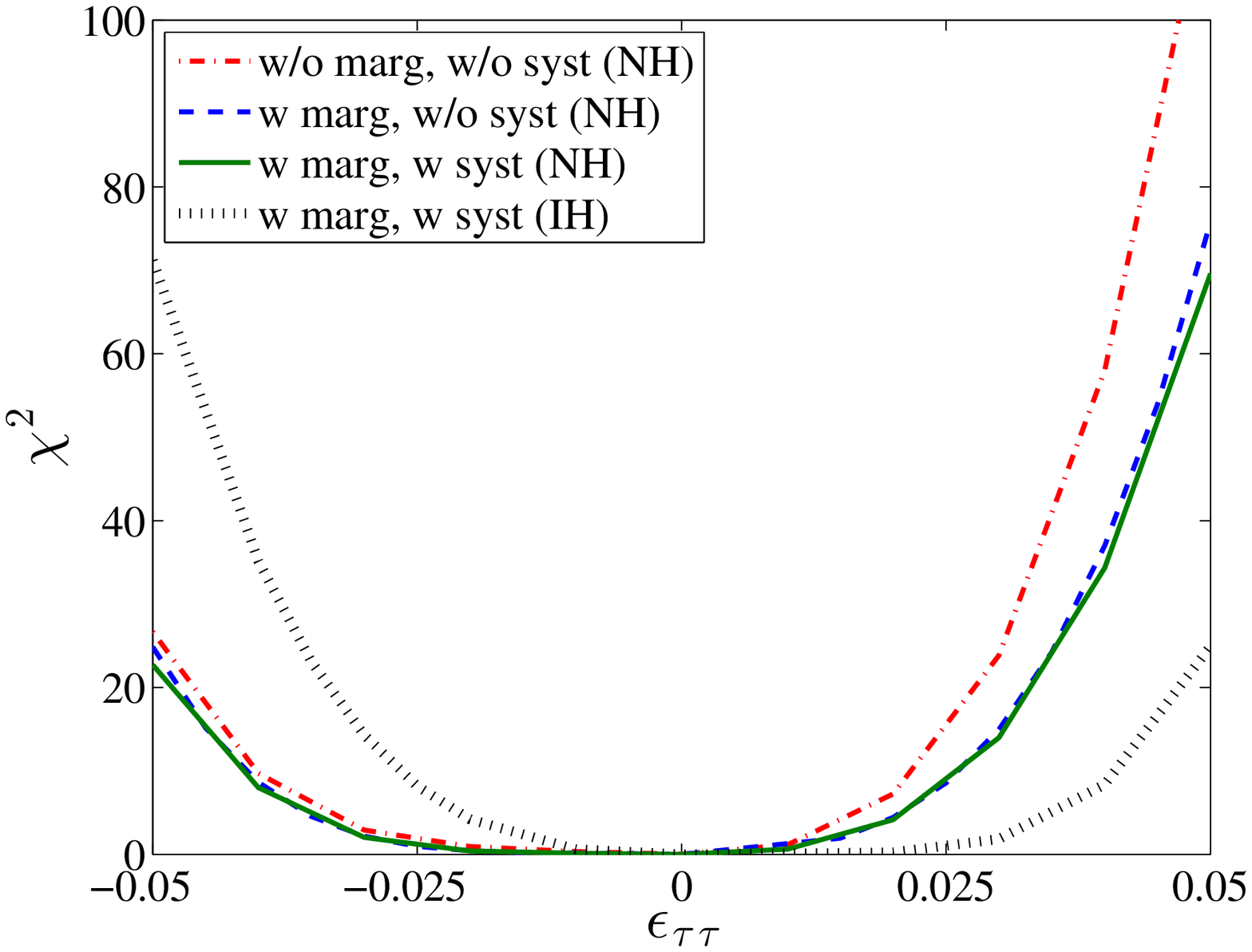}
\caption{The expected $\chi^2$ for NSIs as a function of $\varepsilon_{\mu\tau}$ (left plot) and $\varepsilon_{\tau\tau}$ (right plot) from three years of PINGU data. For normal neutrino mass hierarchy, the red dot-dashed curves are obtained without marginalization and without systematic uncertainties, the blue dashed curves are when marginalization is included but not the systematic uncertainties, while the green solid curves correspond to the case where both marginalization and systematic uncertainties are included. For inverted neutrino mass hierarchy, the black dotted curves include both marginalization and systematic uncertainties.}
\label{fig:diff1}
\end{center}
\end{figure}

In this section, we present results of our statistical analysis of the expected number of events at the PINGU experiment, and the constraints that they could put on the NSI parameters $\varepsilon_{\mu\tau}$ and $\varepsilon_{\tau\tau}$. We take as our ``data" the predicted event rates at PINGU for the following ``true'' values of the neutrino oscillation parameters:
\begin{center}
$\Delta m^2_{21}=7.45\times 10^{-5}$ eV$^2$, \quad
$\Delta m^2_{31}=2.417\times 10^{-3}$ eV$^2$, \\
$\theta_{12} = 33.57^\circ$, \quad
$\theta_{13} = 8.73^\circ$, \quad
$\theta_{23} = 41.9^\circ$, \quad
$\delta_{CP}=0$,
\end{center}
where we have adopted the standard notation for these parameters used in the literature. We next ``fit" these data by minimizing a $\chi^2$ function defined as
\begin{equation}
\chi^2 = \min_{\{\xi_j\}}\sum_{ij}\left[\frac{({N'}^{\rm th}_{ij} - N^{\rm ex}_{ij})^2}{N'^{\rm ex}_{ij}}\right]
+ \sum_{s=1}^k\xi_s^2 \,,
\label{eq:chisq}
\end{equation}
where ${N}^{\rm th}_{ij} $ and $N^{\rm ex}_{ij}$ are the predicted and ``measured" numbers of events in the $i$th zenith angle bin and the $j$th energy bin, respectively, as described in the previous section and the primed ${N'}^{\rm th}_{ij}$ denotes the expected events with the systematic uncertainties taken into account as follows:
\begin{equation}
{N'}^{\rm th}_{ij} = {N}^{\rm th}_{ij}\left(1+\sum_{s=1}^k\pi_{ij}^s \xi_s\right) + {\cal O}(\xi_s^2) \,,
\end{equation}
where $\pi_{ij}^s$ is the uncertainty in the ${ij}$th bin due to the systematic error $s$, and $\xi_s$ is the corresponding pull variable. We have included five systematic uncertainties in our analysis. We take a flux normalization uncertainty of 20~\%, cross-section normalization uncertainty of 10~\%, additional uncorrelated detector uncertainty of 5~\%, zenith angle dependent uncertainty in the flux of 5~\%, and an energy dependent uncertainty in the flux of 5~\%. While the first three systematic errors affect only the normalization of the events, the last two change the energy and zenith angle spectral shape. In our analysis, we have taken ten zenith angle bins of equal width 0.1 in $\cos\theta_z$ between $\cos\theta_z=-1$ to 0, and 27 energy bins, twelve of which have width 2~GeV between 2~GeV and 26~GeV, and the remaining 15 have width 5~GeV between 26~GeV and 101~GeV. 

In Fig.~\ref{fig:diff1}, we present the median $\chi^2$ after three years of running of the PINGU experiment as a function of the NSI parameters $\varepsilon_{\mu\tau}$ (left plot) and $\varepsilon_{\tau\tau}$ (right plot). To generate this plot, we use ``data" which conform to standard oscillations with NSI parameters assumed to be zero. We next fit these ``data" for non-zero values of $\varepsilon_{\mu\tau}$ and $\varepsilon_{\tau\tau}$. The red dot-dashed curves, the blue dashed curves, and the green solid curves are for data corresponding to normal neutrino mass hierarchy, while the black dotted curves correspond to inverted neutrino mass hierarchy. With true normal hierarchy, the red dot-dashed curves show the $\chi^2$ without marginalization over the neutrino oscillation parameters and without including the systematic uncertainties. Switching on the marginalization over the parameters $\Delta m_{31}^2$, $\sin^2\theta_{13}$, and $\sin^2\theta_{23}$ lowers the $\chi^2$ and the  corresponding results are given by the blue dashed curves. We have checked that marginalization over $|\Delta m^2_{31}|$ has the largest impact on the $\chi^2$. When marginalizing, we put priors on the three neutrino oscillation parameters with $1\sigma$ errors as follows: 1~\% for $|\Delta m_{31}^2|$, 2~\% for $\sin^2\theta_{23}$, and 0.005 for $\sin^22\theta_{13}$. We have checked that  marginalizing the $\chi^2$ over $\varepsilon_{\tau\tau}$ in the left plot and $\varepsilon_{\mu\tau}$ in the right plot of Fig.~\ref{fig:diff1} makes negligible difference to the outcome.

Once both marginalization as well as systematic uncertainties are included, we obtain the green solid curves. Note that the systematic uncertainties in this case do not seem to make too much impact on the sensitivity of the experiment to the NSI parameters. This is in contrast to the case of neutrino mass hierarchy determination at PINGU, where the energy and zenith angle dependent uncorrelated errors make a significant reduction of the $\chi^2$. The neutrino mass hierarchy determination crucially depends on the difference between the number of predicted muon events between normal and inverted hierarchies and this is important in the energy range (5--10)~GeV, where it fluctuates fast between energy and zenith angle bins. Therefore, the energy and zenith angle dependent errors can dilute the $\chi^2$ for neutrino mass hierarchy determination. However, for NSIs, the relevant energy ranges in the interval between (10--100)~GeV, where the fluctuation from bin to bin is not significant. This is evident from Figs.~\ref{fig:diffepsmt} and \ref{fig:diffepsboth}. It is noteworthy that detector resolutions do dilute and wash-out to a large extent the NSI contributions at energies lower than 10~GeV, as is obvious by comparing the probability figures with the event figures, as was discussed before. Finally, the black dotted curves are for true inverted hierarchy with both marginalization and systematic uncertainties taken into account. Note that the trend of the $\chi^2$ curves between positive and negative values of $\varepsilon_{\mu\tau}$ and $\varepsilon_{\tau\tau}$ reverse when we change from normal to inverted hierarchy. For normal hierarchy, we have better constraints for negative $\varepsilon_{\mu\tau}$ and positive $\varepsilon_{\tau\tau}$, whereas for inverted hierarchy, we have better constraints for positive $\varepsilon_{\mu\tau}$ and negative~$\varepsilon_{\tau\tau}$.

With three years of data, a median experiment like PINGU is expected to constrain the NSI parameters at the 90~\% C.L.~to
\vspace{-0.5mm}
\begin{align}
-0.0043~(-0.0048) &< \varepsilon_{\mu\tau} < 0.0047~(0.0046) \,, \nonumber\\
-0.03~(-0.016) &< \varepsilon_{\tau\tau} < 0.017~(0.032) \nonumber
\end{align}
for normal (inverted) neutrino mass hierarchy, with the projected uncertainties over standard oscillation parameters taken into account. At the $3\sigma$ C.L., the corresponding bounds are expected to be
\vspace{-5mm}
\begin{align}
-0.0074~(-0.0086) &< \varepsilon_{\mu\tau} < 0.0086~(0.0081) \,, \nonumber\\
-0.04~(-0.025) &< \varepsilon_{\tau\tau} < 0.025~(0.04) \nonumber 
\end{align}
for normal (inverted) neutrino mass hierarchy. We have used the definition where 90~\% C.L.~corresponds to $\Delta \chi^2=2.71$, while $3\sigma$ corresponds to $\Delta \chi^2=9$. Therefore, PINGU with just three years of running could constrain the NSI parameters nearly one order of magnitude better than the current bounds from the Super-Kamiokande experiment. 

\begin{figure}[t]
\begin{center}
\includegraphics[width=0.83\textwidth]{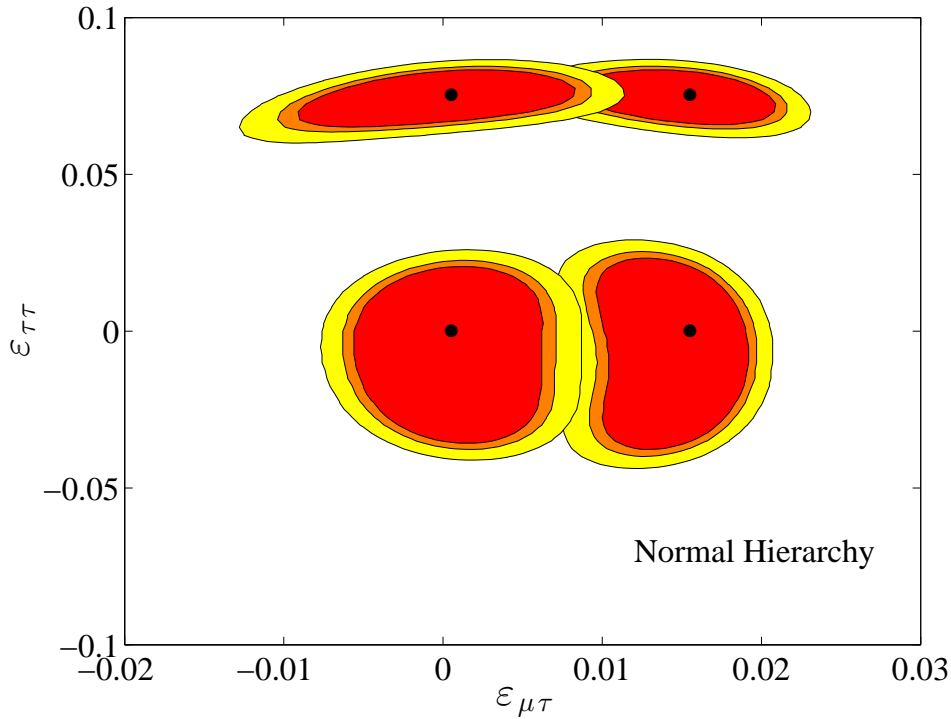}
\caption{The 90~\% (red), 95~\% (orange), and 99~\% (yellow) C.L.~contours in the $\varepsilon_{\mu\tau}$-$\varepsilon_{\tau\tau}$ plane, expected from three years of running of the PINGU experiment for four different choices of the true values for the NSI parameters $(\varepsilon_{\mu\tau},\varepsilon_{\tau\tau})$ of $(0,0)$, $(0.015,0)$, $(0,0.075)$, and $(0.015,0.075)$. The fit is performed by marginalizing over the standard oscillation parameters $\Delta m_{31}^2$, $\sin^2\theta_{13}$, and $\sin^2\theta_{23}$, with priors imposed on them as described in the text. The black dots denote the points at which the ``data'' were considered. The allowed area for a given data set is unique and exists around the (black) point where the data were considered. Normal neutrino mass hierarchy was assumed to be true for all four cases.
\label{fig:cont}}
\end{center}
\end{figure}
 
In Fig.~\ref{fig:cont}, we display the 90~\% (red), 95~\% (orange), and 99~\% (yellow) C.L.~contours in the $\varepsilon_{\mu\tau}$-$\varepsilon_{\tau\tau}$ plane, which are expected from three years of running of PINGU. The four patches in this figure correspond to four different data sets considered for the assumed true values of the NSI parameters $(\varepsilon_{\mu\tau},\varepsilon_{\tau\tau})$ that equal to $(0,0)$, $(0.015,0)$, $(0,0.075)$, and $(0.015,0.075)$. These are shown by black dots in the figure. In addition, we have assumed normal neutrino mass hierarchy for all cases. The fit is marginalized over the standard oscillation parameters $\Delta m_{31}^2$, $\sin^2\theta_{13}$, and $\sin^2\theta_{23}$. The allowed area for a given data set is unique and exists around the (black) point where the data were considered. Note that the shape of the contours changes as we change the assumed allowed set of true values for $(\varepsilon_{\mu\tau},\varepsilon_{\tau\tau})$. The allowed range of $\varepsilon_{\tau\tau}$ decreases as the true value of $\varepsilon_{\tau\tau}$ increases, irrespective of the true value of $\varepsilon_{\mu\tau}$. The change is less marked on the allowed range of $\varepsilon_{\mu\tau}$. However, one still obtains a larger allowed range for $\varepsilon_{\mu\tau}$ if the NSI parameters were non-zero compared to the case when there are no NSIs. This figure shows more expected allowed ranges for $\varepsilon_{\mu\tau}$ and $\varepsilon_{\tau\tau}$ at any given C.L.~than Fig.~\ref{fig:diff1} for two reasons. Firstly, both $\varepsilon_{\mu\tau}$ and $\varepsilon_{\tau\tau}$ are varying in the fit, and secondly, the definition of C.L.~chosen in this figure corresponds to a two-parameter fit ($\Delta \chi^2=9.21$, 5.99, and 4.61 for 99~\% C.L., 95~\% C.L., and 90~\% C.L., respectively). The shape of the contour suggests that in atmospheric neutrino experiments like PINGU, there is very little correlation between the NSI parameters $\varepsilon_{\mu\tau}$ and $\varepsilon_{\tau\tau}$. We have checked that no new features arise in the allowed C.L. contours when inverted hierarchy is assumed to be the true hierarchy. Hence, we do not explicitly show the corresponding contours for the inverted mass hierarchy.

\section{Summary and Conclusions}
\label{sec:s&c}

In this work, we have investigated the impact of the NSI parameters $\varepsilon_{\mu\tau}$ and $|\varepsilon_{\tau\tau} - \varepsilon_{\mu\mu}|$ on the PINGU atmospheric neutrino experiment, which is a planned extension of the already operational IceCube (and DeepCore) experiment(s) that has been successfully observing very high energy neutrinos in the 10~GeV to a few PeV energy range. This experiment has already collected a large body of data on very high energy neutrinos, including the ones coming from the Earth's atmosphere. The PINGU proposal plans to reduce the energy threshold of the detector to less than 5 GeV by placing more strings carrying the optical modules in a small region of the detector. This lower energy threshold would enable the observation of Earth matter effects for atmospheric neutrinos in the (5--10)~GeV range, thereby leading to the determination of the neutrino mass hierarchy. With a multi-megaton effective volume, PINGU could determine the neutrino mass hierarchy to a very high statistical significance within a short time. Prospects of looking for NSIs at PINGU has been discussed in the literature before. In this work, we have further extended these studies by looking at the bounds on NSI parameters from PINGU in the (2--100)~GeV range.

We have presented approximate analytical formulas giving the difference in the muon neutrino survival probability $\Delta P_{\mu\mu}$ in the presence of the NSI parameters $\varepsilon_{\mu\tau}$ and $|\varepsilon_{\tau\tau} - \varepsilon_{\mu\mu}|$ and plotted them in the $h$-$\log E$ plane. The probability difference $\Delta P_{\mu\mu}$ was given for the three-flavor case and the so-called two-flavor hybrid model. A comparison of these approximate results against the exact $\Delta P_{\mu\mu}$ calculated numerically revealed that the approximate three-flavor formula failed to reproduce the correct  $\Delta P_{\mu\mu}$, especially at higher energies and for values of $|\varepsilon_{\tau\tau} - \varepsilon_{\mu\mu}|$ close to the upper bound from Super-Kamiokande. On the other hand, the hybrid model agreed reasonably well with the exact numerical results, modulo the small difference at low energies coming from the fact that a step-function Earth matter density profile was used for the former, while the PREM profile was used for the latter. The impact of the NSI parameters was observed to be largest at $E\sim 25$~GeV. We showed the impact of $\varepsilon_{\mu\tau}$ and $\varepsilon_{\tau\tau}$ (setting $\varepsilon_{\mu\mu}=0$) on $\Delta P_{\mu\mu}$ separately. We also showed the impact of changing the sign of $\varepsilon_{\mu\tau}$ and $\varepsilon_{\tau\tau}$ on $\Delta P_{\mu\mu}$ as well as the impact of the neutrino mass hierarchy.

We have calculated the events due to atmospheric neutrinos at PINGU and plotted the difference between the number of events given by standard oscillations and oscillations in presence of NSI parameters in the $\cos\theta_z$-$\log E$ plane. The event plots followed the trend expected from the probability plots, modulo the smearing out of the energy and zenith angle fluctuations due to the finite energy and zenith angle resolution of the detector incorporated in our calculations. Next, we have performed an estimate of the sensitivity of PINGU to NSI parameters by defining a $\chi^2$ function, taking into account the systematic uncertainties in the atmospheric neutrino fluxes. In the case that there were no NSIs in the neutrino sector, with three years of data, PINGU should be able to constrain the NSI parameters at the 90~\% C.L.~to
\vspace{-5mm}
\begin{align}
-0.0043~(-0.0048) &< \varepsilon_{\mu\tau} < 0.0047~(0.0046) \,, \nonumber\\
-0.03~(-0.016) &< \varepsilon_{\tau\tau} < 0.017~(0.032) \nonumber
\end{align}
for normal (inverted) neutrino mass hierarchy, after marginalizing the $\chi^2$ function over the projected uncertainties of the standard neutrino oscillation parameters. Finally, we have showed the expected C.L.~allowed contours in the $\varepsilon_{\mu\tau}$-$\varepsilon_{\tau\tau}$ plane with three years of PINGU data if the NSIs exist in Nature. The contour also implies that there is basically no correlation between $\varepsilon_{\mu\tau}$ and $\varepsilon_{\tau\tau}$.

\begin{acknowledgments}
This work was supported by the Swedish Research Council (Vetenskapsr{\aa}det), contract No.~621-2011-3985 (T.O.). S.C.~acknowledges support from the Neutrino Project under the XII plan of Harish-Chandra Research Institute and partial support from the European Union FP7 ITN INVISIBLES (Marie Curie Actions, PITN-GA-2011-289442). T.O.~thanks the KITP in Santa Barbara for kind hospitality. This research was supported in part by the National Science Foundation under Grant No.~NSF PHY11-25915.
\end{acknowledgments}

\end{document}